\documentclass{aa}
\usepackage{epsf}

\newcommand{\xt}[1]{\mbox{$\times 10^{#1}$}} 
 
\def\nt#1{\vtop{\footnotesize\hsize=\columnwidth\leavevmode#1\hspace*{\fill}}}
\def\ntd#1{\vtop{\footnotesize\hsize=\textwidth\leavevmode#1\hspace*{\fill}}}

\begin{document}
\title{Shock emission in the bipolar post-AGB star IRAS~16594$-$4656\thanks{Based on observations collected at the
European Southern Observatory, Chile (proposal no.\ 61.C-0567).}}
\subtitle{}
\author{G. C. Van de Steene\inst{1} 
\and P. A. M. van Hoof \inst{2,3}}

\institute{ Royal Observatory of Belgium, Ringlaan 3, 1180 Brussels, Belgium
\and APS Division, Physics Dept., Queen's University of Belfast, BT7 1NN, Northern Ireland
\and Canadian Institute for Theoretical Astrophysics, McLennan Labs, University of Toronto, 60 St. George St., Toronto, ON M5S~3H8, Canada
}

\offprints{G. C. Van de Steene, \email{gsteene@oma.be}}

\authorrunning{Van de Steene \& van Hoof}
\titlerunning{Shock emission in IRAS~16594$-$4656}

\date{Received / Accepted}

\abstract{ In this paper we study the near-infrared emission spectrum of
  IRAS~16594$-$4656, a bipolar post-AGB star with spectral type B7 and
  no observed ionization. Using optical and near-infrared photometry we
  determined the total extinction towards this object to be
  $A_V=7.5\pm0.4$~mag and derived a distance of $2.2\pm0.4$~kpc,
  assuming a luminosity of $10^4$~L$_\odot$. The near-infrared spectrum
  shows strong H$_2$ emission lines and some typical metastable shock
  excited lines such as [Fe\,{\sc ii}] 1.257~\& 1.644~$\mu$m. We
  determined the rotational and vibrational excitation temperatures, as
  well as the ortho-to-para ratio of the molecular hydrogen. Based on
  these we argue that the H$_2$ emission is mainly collisionally
  excited. Line ratios indicate that the H$_2$ emission originates in a
  $\sim$25~km\,s$^{-1}$ C-type shock. On the other hand, the metastable
  lines, and especially the [Fe\,{\sc ii}] emission lines, indicate the
  presence of a $\sim$75~km\,s$^{-1}$ J-type shock. Hence we postulate
  that the H$_2$ emission originates where the stellar wind (with an
  observed terminal velocity of $\sim$126~km\,s$^{-1}$) is funneled
  through an equatorial density enhancement, impinging almost
  tangentially upon the circumstellar material. The [Fe\,{\sc ii}]
  emission either occurs along the walls of the bipolar lobes where the
  transverse shock velocity would be higher, or could originate much closer
  to the central star in shocks in the post-AGB wind itself, or possibly
  even an accretion disk. Further high resolution near-infrared spectra
  are currently being obtained to confirm the proposed geometry and
  kinematics.

   }

\maketitle

\section{Introduction}

Post-AGB stars represent an important transition phase in the
evolution of low and intermediate-mass stars, between the asymptotic
giant branch (AGB) and the planetary nebula (PN) phases. During this
period, the detached circumstellar envelope of gas and dust is expanding
away from the star.  Meanwhile the star itself is increasing in
temperature at about constant luminosity.  This phase lasts a few
thousand years depending upon the star's core mass.  When the
temperature is high enough and the star photo-ionizes the nebula,
it has entered the PN phase.

In spite of extensive study, the evolution from the AGB toward the PN
stage is still poorly understood.  The drastic changes observed in
circumstellar structure and kinematics are particularly puzzling.
During late AGB or early post-AGB evolutionary stages, the
geometry of the circumstellar material changes from more or less spherically symmetric to axially
symmetric, with the result that most PNe exhibit axisymmetric
structures, ranging from elliptical to bipolar.  Bipolar PNe are very
likely to possess molecular envelopes that are readily detectable in
the near-infrared ro-vibrational lines of H$_2$ (Kastner et al. 
\cite{Kastner96}).
The available data suggest that the onset of near-infrared H$_2$
emission in PNe can be traced back to the pre-planetary nebula (PPN)
phase but not back to the AGB phase of evolution (Weintraub et al.
\cite{Weintraub98}). These observations suggest that further studies of H$_2$
emission from PPNe may offer insight into the transition from AGB star
to PN and from spherical to axisymmetric mass loss.

The study of transition objects showing H$_2$ emission at an early
stage is crucial for understanding the hydrodynamic processes shaping
the nebulae. The H$_2$ lines can reveal details about the physical conditions
in the shocks associated with these hydrodynamic processes and thereby
help constrain models of the interaction of the central star with the AGB remnant.
Only 13
post-AGB stars with H$_2$ emission have been detected in molecular hydrogen
(Garc\'\i a-Hern\'andez et al. \cite{GarciaH02}). 
RAFGL 2688 (the Egg Nebula) and RAFGL 618
are perhaps the best-studied examples of such transition objects and
the only two whose H$_2$ emission has been extensively studied.  Both
objects display molecule-rich circumstellar envelopes, carbon-rich
circumstellar chemistries, and dusty, bipolar reflection nebulosities.
Whereas RAFGL 2688 harbors an F-type central star and is not known to
contain an H\,{\sc ii} region, RAFGL 618 possesses a B-type central star and
does contain a compact H\,{\sc ii} region, suggesting that the latter nebula has evolved
further toward the PN stage.  In this paper we study in detail the
excitation mechanism of the H$_2$ emission in the PPN
IRAS~16594$-$4656.

IRAS~16594$-$4656 is classified as a post-AGB star for several
reasons.  In the IRAS color-color diagram it has the colors of a
planetary nebula (PN) (Van de Steene \& Pottasch \cite{VdSteene93}),
but it has never been detected in the radio (Van de Steene et al.\ 
\cite{VdSteene00a}).  It has a large infrared excess due to dust with
a color temperature of 173~K. It displays a double-peaked spectral
energy distribution, with the peak in the mid-infrared much brighter
than the peak in the near-infrared (Van de Steene et al.\ 
\cite{VdSteene00a}).  It possesses a CO envelope with an expansion
velocity of at least 16~km\,s$^{-1}$ (Loup et al.\ \cite{Loup90}).  The
chemistry of IRAS~16594$-$4656 appears to be carbon-rich. This is
based on the detection of unidentified IR emission features at 3.3,
6.2, 7.7, 8.6, 11.3, 12.6, and 13.4~$\mu$m, as well as the 21~$\mu$m
feature (Garc\'\i a-Lario et al.\ \cite{GarciaL99}), all commonly
associated with a carbon-rich chemistry. However, tentative detections
of crystalline silicate features (mostly pyroxenes) were also reported
by Garc\'\i a-Lario et al.\ (\cite{GarciaL99}). The latter are
associated with an oxygen-rich chemistry, which could point to the
fact that IRAS~16594$-$4656 only turned carbon-rich during one of the
most recent thermal pulses. 

The optical spectrum of IRAS~16594$-$4656 shows a spectral type B7
with significant reddening. The Balmer lines of hydrogen are in
emission, together with a weak [O\,{\sc i}] emission at 6300~\AA.  The
H$\alpha$ emission has a P-Cygni type profile indicative of a stellar
wind with a terminal velocity of approximately 126~km\,s$^{-1}$ (Van
de Steene et al.\ \cite{VdSteene00b}).  

In the optical the nebulosity is dominated by scattered light, not
emission lines as in the case of planetary nebulae (Hrivnak et al.\ 
\cite{Hrivnak99}).  The {\it HST} image of IRAS~16594$-$4656 shows a
bipolar morphology (Hrivnak et al.\ \cite{Hrivnak99}), with the
southwestern lobe tilted towards us at an intermediate orientation (Su
et al. \cite{su01}), as well as concentric arcs (Hrivnak et al.\ 
\cite{Hrivnak01}).  Su et al.\ (\cite{su00}) noted that in addition to
the centro-symmetric polarimetric patterns, point-symmetric patterns
are also seen in IRAS~16594$-$4656. Such patterns provide strong,
independent evidence for the presence of a circumstellar disk or torus
(hereafter referred to as an equatorial density enhancement or EDE).

Spectroscopic studies of H$_2$ are particularly useful because the
intensity ratios of H$_2$ lines arising from different excited levels
provide diagnostics of the excitation mechanism and physical
conditions.  There are two likely excitation mechanisms for H$_2$ in
post-AGB stars: UV pumping (``fluorescence'') by stellar photons, and
thermal (i.e., collisional) excitation by the gas.  It is possible
that both mechanisms play a role within a single post-AGB star.  All
spectroscopic (and imaging) studies of post-AGB stars to date have
concentrated only on the strong H$_2$ (1,0)\,S(1) 2.121~$\mu$m and
H$_2$ (2,1)\,S(1) 2.247~$\mu$m spectral lines.  Their ratio can in
principle distinguish between the two excitation mechanisms.  However,
its utility as a diagnostic breaks down at high densities, when
collisions modify the populations of the lowest vibrational levels
($v=1,2$), driving them to their thermal values, even when the gas is
fluorescently excited.  Emission lines from higher vibrational levels
are less affected, and can still distinguish gas in which UV photons
are responsible for the H$_2$ excitation.  Furthermore time-dependent
effects might enhance the intensity of UV-pumped emission soon after
the onset of far-UV irradiation, enabling fluorescent H$_2$ emission
to rival the high surface brightness typical of shocked gas.  This
makes it particularly important to study more than a handful of
post-AGB stars, over a large spectral range, in order to achieve a
broad, balanced picture of their molecular components.

For the first time we present a full {\it J}-, {\it H}-, {\it K}-band
spectrum of a post-AGB star. The observations are presented in
Sect.~\ref{observations}. The extinction and distance are derived in
Sect.~\ref{distance} and the infrared spectrum is discussed in
Sect.~\ref{lines}.  Many H$_2$ emission lines were detected and we
consider them all to discuss the excitation mechanism in
Sect.~\ref{excitation}.  The results are discussed in Sect.~\ref{disc}
and final conclusions are given in Sect.~\ref{concl}.

\section{Observations}
\label{observations}

\begin{figure}
\caption{  The HST mage of IRAS~16594$-$4656 through filter F814W (Hrivnak et
  al.\ \cite{Hrivnak99}) with the slit position indicated by the grey
  lines. North is to the top and East to the left.}
\mbox{\epsfxsize=0.95\columnwidth\epsfbox[36 184 577 608]{IRAS1659slit.ps}}
\label{image}
\end{figure}

\begin{figure}
\mbox{\epsfxsize=0.95\columnwidth\epsfbox[47 300 510 653]{h2plot2.ps}}
\mbox{\epsfxsize=0.95\columnwidth\epsfbox[47 300 510 653]{h2plot3.ps}}
\mbox{\epsfxsize=0.95\columnwidth\epsfbox[47 300 510 653]{h2plot1.ps}}
\caption{The H$_2$ spectrum of IRAS~16594$-$4656.}
\label{spec:i}
\end{figure}

\begin{figure}
\mbox{\epsfxsize=0.95\columnwidth\epsfbox[47 300 510 653]{h2plot4.ps}}
\mbox{\epsfxsize=0.95\columnwidth\epsfbox[47 300 510 653]{h2plot5.ps}}
\mbox{\epsfxsize=0.95\columnwidth\epsfbox[47 300 510 653]{h2plot6.ps}}
\caption{The other emission lines of IRAS~16594$-$4656.}
\label{spec:ii}
\end{figure}

The near-infrared spectrum of IRAS~16594$-$4656 was obtained with {\it
  SOFI} on the {\it NTT} telescope at La Silla (ESO) in June 1998.
The slit width was 1\arcsec\ and the slit orientation almost
east-west. An image with the slit position indicated is presented in
Fig. \ref{image}.  The {\it SOFI} 
 pixel scale is 0\farcs29 per pixel.  Between 0.95
and 1.65 $\mu$m the spectrum was obtained with the GB grism and has a
spectral resolution of 7~\AA\ per pixel.  Between 1.53 and 2.52
$\mu$m the spectrum was obtained with the GR grism and has a
resolution of 10.2~\AA\ per pixel.  The seeing was 0\farcs7. The red
spectrum is an average of 6 cycles of 10 DITs of 1.5~sec and the blue
spectrum an average of 6 cycles of 10 DITs of 2~sec. The airmass was 1.45.

The spectra were reduced with an adapted version of the ESO package
{\sc eclipse} and further processed and analyzed in {\sc iraf}. A
selection of 3 standard stars, observed throughout the night, was used
to construct the response curve.  To obtain absolute flux tables for
the standard stars, we constructed theoretical spectra for the
standard stars based on NextGen atmosphere models produced with the
stellar atmosphere code {\sc phoenix} (Hauschildt et al.\ 
\cite{hauschildt99}), which were subsequently convolved to the {\it
  SOFI} resolution and wavelength grid.  Next we included an
extinction component and scaled the spectra to match the observed
{\it J}-, {\it H}-, and {\it K}-band flux values. These spectra were
compared with the observed ones to derive the average response curve.
The object spectra were calibrated with this smoothed response curve.
Comparison of the observed, calibrated spectra of the standard stars
and the theoretical spectra provided us with the spectrum of the
atmosphere, which was subsequently used to correct for telluric
absorption.

\begin{table}
\caption{The observed and intrinsic magnitudes of IRAS~16594$-$4656. The
intrinsic magnitudes are arbitrarily scaled to {\it V} = 0~mag. }
\label{photom}
\small
\begin{tabular}{lrrrrrrr}
\hline
\hline
 & \it B$^a$ & \it V$^a$ & \it R$_{\rm C}^a$ & \it I$_{\rm C}^a$ & \it J$^b$ & \it H$^b$ & \it K$^b$  \\
& mag & mag & mag & mag & mag & mag & mag \\
\hline
$m_{\rm obs}$\hspace*{-2mm}  & 16.31 & 14.60 & 13.27 & 12.00 & 9.73 & 8.85 & 8.20  \\
$m_{\rm intr}$\hspace*{-2mm} & $-$0.09 & 0.00 & 0.01 & 0.06 & 0.16 & 0.18 & 0.23  \\
\hline
\end{tabular}
\nt{$^a$ Hrivnak et al.\ (\cite{Hrivnak99})}
\nt{$^b$ Van de Steene et al. (\cite{VdSteene00a})}
\end{table}

\begin{table*}
\caption{The emission lines detected in the spectrum of IRAS~16594$-$4656. 
The observed flux values $F_{\rm obs}$ in the 5$^{\rm th}$ column are not corrected for extinction, 
but the $F_{\rm cor}$ values in the last column are corrected for extinction.}
\begin{flushleft}
\begin{tabular}{lllllll}
\hline
Line     & $\lambda_{\rm lab}$ & $\lambda_{\rm obs}$ & Continuum & $F_{\rm obs}$      &  FWHM   & $F_{\rm cor}$ \\
         &   $\mu$m  & $\mu$m & erg cm$^{-2}$ s$^{-1}$ \AA$^{-1}$ & erg cm$^{-2}$ s$^{-1}$ &  $\mu$m  &  erg cm$^{-2}$ s$^{-1}$ \\
\hline
(1,0)\,S(3) & 1.9570 & 1.9575 & 1.91($-$14)$^a$&1.66($-$13)& 3.13($-$3) & 4.88($-$13) \\
(1,0)\,S(2) & 2.0332 & 2.0336 & 1.93($-$14) & 1.07($-$13) & 3.69($-$3)  & 2.88($-$13) \\
(1,0)\,S(1) & 2.1213 & 2.1216 & 1.88($-$14) & 3.17($-$13) & 3.36($-$3)  & 7.87($-$13) \\
(1,0)\,S(0) & 2.2227 & 2.2229 & 1.71($-$14) & 9.15($-$14) & 3.61($-$3)  & 1.95($-$13) \\
(2,1)\,S(1) & 2.2471 & 2.2478 & 1.67($-$14) & 2.62($-$14):& 4.01($-$3): & 4.87($-$14): \\
(1,0)\,Q(1) & 2.4059 & 2.4061 & 1.47($-$14) & 3.71($-$13) & 3.65($-$3)  & 7.78($-$13) \\
(1,0)\,Q(2) & 2.4128 & 2.4130 & 1.43($-$14) & 1.32($-$13) & 3.88($-$3)  & 2.63($-$13) \\
(1,0)\,Q(3) & 2.4231 & 2.4232 & 1.44($-$14) & 2.91($-$13) & 3.54($-$3)  & 5.94($-$13) \\
(1,0)\,Q(4) & 2.4368 & 2.4370 & 1.42($-$14) & 9.66($-$14) & 3.04($-$3)  & 1.86($-$13) \\
(1,0)\,Q(5) & 2.4541 & 2.4544 & 1.39($-$14) & 9.91($-$14) & 3.04($-$3)  & 2.02($-$13) \\
(1,0)\,Q(7) & 2.4993 & 2.4989 & 1.22($-$14) & 8.53($-$14):& 4.25($-$3): & 1.75($-$13): \\
\hline
[C\,{\sc i}]   & {\it 0.9844$^b$}   & 0.9849 & 1.88($-$14) &1.16($-$13) & 2.71($-$3): &  2.98($-$12) \\ \relax
[Fe\,{\sc ii}] &     1.2567     & 1.2567 & 2.76($-$14) &  3.88($-$14): & 2.12($-$3) & 3.23($-$13):\\ \relax
Pa$\beta$      &     1.2818     & 1.2821 & 2.73($-$14) &  2.34($-$13)  & 2.09($-$3) & 1.88($-$12) \\ \relax
[Fe\,{\sc ii}] &     1.6435     & 1.6437 & 2.55($-$14) &  3.66($-$14)  & 2.17($-$3) & 8.70($-$14) \\ \relax
Br$\gamma$     &     2.1655     & 2.1662 & 1.80($-$14) &  4.52($-$14) & 2.91($-$3) & 9.42($-$14) \\\hline
\end{tabular}
\ntd{$^a$ Entries such as ($-$14) stand for \xt{-14}.}
\ntd{$^b$ This line is a blend of the [C\,{\sc i}] 0.9824 \& 0.9850~$\mu$m lines.}
\end{flushleft}
\end{table*}

\section{Extinction and distance}
\label{distance}

IRAS~16594$-$4656 lies very close to the Galactic plane ($b =
-3$\fdg3).  As such, it is expected to be significantly reddened.
Because the reddening must be accounted for in the analysis of the
emission lines and the continuum, we begin this section with a
determination of the reddening. We used {\it BVR$_{\rm C}$I$_{\rm C}$}
photometry taken from Hrivnak et al.\ (\cite{Hrivnak99}) and {\it JHK}
photometry taken from Van de Steene et al.\ (\cite{VdSteene00a}). We
excluded the observed {\it L}-band magnitude from our analysis since it is
apparently contaminated by emission from either atomic lines or hot dust.
The spectral type of the central star is B7 (Van de Steene et al.\ 
\cite{VdSteene00b}) and we assumed that the star has a low surface
gravity. The intrinsic colors of the central star were derived from an
{\sc atlas9} stellar atmosphere model (Kurucz \cite{kurucz94}) with
$T_{\rm eff} = 12000$~K and $\log [g/({\rm cm\,s^{-2}})] = 2.00$.  The
data are summarized in Table~\ref{photom}.

We determined the total extinction (i.e., interstellar and
circumstellar extinction combined) for each combination of observed
photometric bands (excluding {\it V$-$R$_{\rm C}$}) such that the
dereddened color would match the intrinsic color. We then averaged all
the measurements of $A_V$ and determined the standard deviation. We
used the extinction law from Cardelli et al.\ (\cite{cardelli89}) and
determined $R_V$ by minimizing the standard deviation. The resulting
values are: $A_V = 7.5 \pm 0.4$~mag, with $R_V = 4.2$. Note that the
uncertainty quoted here is the uncertainty in the average value, not
the spread of the individual measurements. The galactic extinction
estimate at the position of IRAS~16594$-$4656 is $A_V=6.6$~mag
according to Schlegel et al.\ (\cite{Schlegel98}), who assume an
$R_V=3.1$ extinction curve. This value is in good agreement with our
determination. The extinction
value is considerably higher than the estimate $A_V \approx 3.2$~mag
given by Hrivnak et al.\ (\cite{Hrivnak99}).  The latter estimate pertains
to the interstellar extinction component alone at an assumed distance of
more than 1.3~kpc, and is based on the galactic extinction study by Neckel \& Klare
(\cite{nk80}).

Visual inspection of the Digitized Sky Survey image taken in the
direction of IRAS~16594$-$4656 reveals that the extinction is very
patchy in the neighborhood of this object. Hence the use of
estimates from galactic extinction studies should be considered very
uncertain. On the other hand, some of the difference between the two extinction values
could be due to significant circumstellar extinction, which would be
supported by the large infrared excess of this object. However it is
unclear how much of the total extinction is of circumstellar origin.

With this value for the extinction and the flux calibration from the Kurucz
model, we can determine that the distance to IRAS~16594$-$4656 is
$(2.2\pm0.4)\,L_4^{1/2}$~kpc, where $L_4$ is the luminosity of
IRAS~16594$-$4656 in units of $10^4$~L$_\odot$. This value is in good
agreement with the distance of $2.5\,L_4^{1/2}$~kpc derived by Su et al.\
(\cite{su01}), based on the dereddened bolometric flux of IRAS~16594$-$4656.
The latter value may have been somewhat overestimated due to stellar radiation
escaping through the lobes. The distance is also somewhat sensitive to
uncertainties in the interstellar extinction estimate, although this effect is
mild since most of the bolometric flux is emitted in the mid-infrared.

\section{Infrared lines}
\label{lines}

\subsection{Atomic hydrogen}
  
The Br$\gamma$ line is in emission, but doesn't show a P-Cygni profile
as we observed in the H$\alpha$ line (Van de Steene et al.\ 
\cite{VdSteene00b}).  The Br$\gamma$ flux detected in the {\it SOFI}
spectrum is approximately 40\% lower than measured from the
higher resolution {\it IRSPEC} spectra obtained in 1995 by Van de
Steene et al.\ (\cite{VdSteene00a}) and in 1994 by Garc\'\i
a-Hern\'andez et al.\ (\cite{GarciaH02}).  For the {\it IRSPEC}
observations the slit had a width of 4\farcs5 and was oriented N-S and E-W,
respectively, while in our {\it SOFI} observations the slit
had a width of only 1\arcsec\ and was oriented E-W.  Based on the continuum
flux, both calibrations agree. This could indicate that we missed some line emission
due to the narrow slit width in the {\it SOFI} observations.
However, this is not a very plausible explanation. The Br$\gamma$ emission is
formed in the post-AGB wind, very close to the central star. The emission
region is much smaller than 1\arcsec\ in diameter, so if we missed Br$\gamma$
emission, we would also have missed continuum emission from the central star.
Reflection also seems a very implausible explanation, since it is not expected
that dust scattering in the infrared would be efficient enough to cause a 40\%
discrepancy in the observed fluxes. This leaves as the only alternative
explanation that the strength of the Br$\gamma$ emission became less between
1994/5 and 1998. We have seen spectral variability in other, similar objects
in our sample, so this possibility cannot be ruled out. Further observations
will be needed to settle this point.

The Pa$\beta$ line has been observed for the first time in this
spectrum, and is shown in Fig~\ref{spec:ii}.

\subsection{Molecular hydrogen}

The observed H$_2$ lines are shown in Fig.~\ref{spec:i}.
The strongest H$_2$ emission is seen from (1,0)\,Q(1), (1,0)\,S(1), and
(1,0)\,Q(3).  The weakest line is (2,1)\,S(1) and no other lines with
$v=2$ or higher are detected. 

Compared with the {\it IRSPEC} value given by Garc\'\i a-Hern\'andez
et al. (\cite{GarciaH02}), we observe $\sim$50\% less flux for the
H$_2$(1,0)\,S(1) line.  The flux of the faint H$_2$(2,1)\,S(1) line
however agrees for both observations (our detection has a large error
margin though). This discrepancy could either be due to flux missed
outside the {\it SOFI} slit, or to variability in the post-AGB wind.
See also the discussion in the previous subsection and
Sect.~\ref{disc}.

\subsection{Carbon}

The [C\,{\sc i}] 9824~\& 9850~\AA\ lines are metastable lines often
observed in shocks (Hollenbach \& McKee \cite{hollenbach89}).
Neutral carbon is readily ionized, so the existence of [C\,{\sc i}]
requires a region sheltered from ionizing radiation, or indicates
the lack of it.

Both [C\,{\sc i}] 9824~\AA\ and 9850~\AA\ come from the same upper
level and have a fixed intensity ratio of approximately 1:3. The fact
that [C\,{\sc i}] 9824~\AA\ is not clearly detected in our spectrum is
most likely due to telluric absorption.

\subsection{Iron}
  
[Fe\,{\sc ii}] lines are known to be good tracers of astrophysical
shocks in supernovae, Herbig-Haro objects, PNe, etc. [Fe\,{\sc ii}] lines
are very sensitive to the high densities and temperatures in the
shocked gas.  The excitation temperature of the upper term of the IR
multiplets of [Fe\,{\sc ii}], which fall in atmospheric windows at 1.2
and 1.6~$\mu$m, is 2-3 times lower than of optical lines and the
critical densities of these [Fe\,{\sc ii}] lines are $\sim$10 times
higher.

The [Fe\,{\sc ii}] $a\,^4\!F$--$a\,^4\!D$ 1.644~$\mu$m over Br$\gamma$
intensity ratio is often used as an indicator of shock excitation.
The ratio expected for shock-excited gas is much larger than 1, but
the [Fe\,{\sc ii}] 1.644~$\mu$m/Br$\gamma$ ratio expected for
radiatively excited gas is only approximately 0.06 (Graham et al.\ 
\cite{Graham87}). The ratio which we obtain for the dereddened line
ratios is 0.9, which is much larger than the value typically found
in H\,{\sc ii} regions.

\section{Excitation of molecular hydrogen}
\label{excitation}

The excitation of molecular hydrogen has been the subject of several
theoretical investigations which are reviewed by Sternberg
(\cite{sternberg90}) and Burton (\cite{burton92}).  Two principal
sources of H$_2$ emission are (1) photodissociation regions (PDRs),
where the molecules are vibrationally excited by far-ultraviolet (FUV)
pumping or collisionally excited in gas heated by FUV radiation; and
(2) shocked regions, in which the hydrogen molecules are collisionally
excited in hot gas behind the shock front.  In post-AGB stars both
forms of excitation may occur. Observations of the rich H$_2$
ro-vibrational spectrum yield line ratios that are valuable probes of
the physical conditions within the emitting source. These allow us
to determine the excitation mechanism.

Garc\'\i a-Hern\'andez et al. (\cite{GarciaH02}) concluded that
fluorescent H$_2$ emission becomes active when the central star
reaches a temperature that corresponds to spectral type A.
Shock excited H$_2$ emission was only detected in objects with a
marked bipolar morphology, sometimes at an early stage during the
post-AGB phase. Sometimes this emission was localized in the waist and/or in
specific regions of the bipolar lobes, where a physical interaction
exists between fast- and slow-moving material. In this section
we investigate the excitation mechanism 
of H$_2$ in IRAS~16594-4656.

\subsection{Rotational and vibrational temperatures of H$_2$}

Because H$_2$ is formed from two identical particles, only quadrupole
transitions are allowed, and transition probabilities are very low. As
a consequence, the radiative lifetimes of the levels are long and all levels
will be populated thermally within a given vibrational state.  If the
relative strengths for a few of the transitions can be measured, then
this information can be used to derive the rotation temperature for
that vibrational state. If collisions alone excite molecular hydrogen,
then populations in different vibrational states will also be in thermal
equilibrium and all levels can be described
by the same temperature. However, if the excitation mechanism includes
a fluorescent component, higher vibrational states will be
overpopulated, thus indicating a vibrational temperature higher than the
rotational value.  Hence, thermal excitation will give similar rotational and
vibrational temperatures, while fluorescence is characterized by a
high vibrational temperature and a low rotational temperature (Rudy et
al.\ \cite{rudy01}).

Fig.~\ref{colden} provides a graphical determination of the temperature and
excitation mechanism of the molecular hydrogen in IRAS~16594$-$4656 in the
standard manner (see, e.g., Martini et al.\ \cite{Martini97}). The $y$-axis
shows the log of the relative column density of molecular hydrogen in a given
state, $N(v,J)$, divided by the statistical weight, $g(J)$. These were derived
by converting the reddening corrected line fluxes $F_{\rm cor}$ into column densities
for the upper levels using the formula: 
\[ N(v,J) = \frac{F_{\rm cor} \lambda}{\Omega h c A_{\rm ki}}. \]
Here all symbols have their usual meaning and $\Omega$ is the solid angle
covered by the observations. The transition probabilities $A_{\rm ki}$ 
were taken
from Martin et al.\ (\cite{martin96}). Since $\Omega$ is unknown for
IRAS~16594$-$4656, we present the data as ratios to $N(1,3)/g(3)$. 
The $x$-axis gives the temperature of the upper level of a given transition.

Comparison of the population of upper levels with different rotational quantum numbers
($J$-values), but identical vibrational quantum numbers ($v$-values), provides
an estimate for the rotational temperature.  The temperature can be
found directly from the slope of a straight line fitted to the data
points.  We have made a least-squares fit using the formula
\[ \ln y - \ln y_0 = a + bx, \]
with $x = E(v,J)/k$, $y = N(v,J)/g(J)$ and $y_0 = N(1,3)/g(3)$.  
The rotational
temperature is simply given by $T_{\rm r} = -1 / b$.  
The results are:
$a = 4.90 \pm 0.30$, $b = (-6.93 \pm 0.41)\xt{-4}$. 
Hence, the rotational temperature
$T_{\rm r} = 1440 \pm 80$~K.

The vibrational temperature is measured from the slope of a line
passing through data points with different vibrational quantum numbers but the
same rotational quantum number in the ${\ln[N(v,J)/g(J)],T}$-plane.
We determined $T_{\rm vib} = 1820 \pm 240$~K.

The fact that the two values differ only by 1.6~$\sigma$ is consistent
with the assumption that fluorescence doesn't contribute strongly to
the excitation and that in IRAS~16594$-$4656 
the H$_2$ excitation is mainly collisional.

\subsection{The ortho-to-para ratio}

The ortho-to-para ratio of molecular hydrogen is the ratio of the total
column density of ortho-H$_{2}$ (all odd $J$ states) to para-H$_2$ (all even
$J$ states).

If the ro-vibrational states are predominantly populated by
collisional processes, the ortho-to-para ratio will be very close to 3. On the other
hand, if FUV fluorescence plays an important role in populating the
levels, the ortho-to-para ratio will be significantly smaller
than 3 (Martini et al.\ \cite{Martini97}). 
Hence the ortho-to-para ratio can be used as a diagnostic to
determine whether FUV fluorescence plays a role in exciting the
ro-vibrational states.

To calculate the ortho-to-para ratio from our observed line fluxes, we
used Eq.~5 from Hoban et al.\ (\cite{hoban91}) with the transition
probabilities and level energies taken from Martin et al.\
(\cite{martin96}). Since we observed many more lines than they did, we
chose to calculate the ortho-to-para ratio from each combination of
ortho and para states that was observed in our spectrum. We also
assigned an uncertainty to each ratio, based on the estimated
uncertainties in the line fluxes and calibration. We then calculated
the weighted average of all ratios, resulting in an ortho-to-para
ratio of 2.77 $\pm$ 0.19.  Note that the uncertainty quoted here is
the uncertainty in the average value, not the spread of the individual
measurements. This value is in good agreement with the expected ratio
of 3 for collisionally excited molecular hydrogen.

\begin{figure}
\mbox{\epsfxsize=0.95\columnwidth\epsfbox[47 318 510 653]{h2lines.ps}}
\caption{The column densities for the various observed ro-vibrational
levels of H$_2$. The best-fit line based on a rotational
temperature of $T_{\rm r} = 1440$~K is shown as well.}
\label{colden}
\end{figure}

\subsection{The (1,0)\,S(1)/(2,1)\,S(1) ratio}

A diagnostic that has been commonly used to discriminate between
shocks and FUV-pumped fluorescence, is the (1,0)\,S(1)/(2,1)\,S(1)
ratio.  In shocks this ratio is typically $\sim$10 (e.g., Shull \&
Hollenbach \cite{shull78}, and references therein) while for pure
radiative fluorescence it is 1.8 (Black \& Dalgarno \cite{black76}).
This ratio is essentially based on the same physics as the comparison
between the rotational and vibrational temperatures, but is a useful 
diagnostic when no other H$_2$ lines have been observed.  The
dereddened value we obtain for this ratio is 16.2, thus indicating
shock excitation.

\subsection{Thermal excitation}

Based on the evidence presented in the previous subsections, 
we can state that the H$_2$ emission
is collisionally excited by other atoms or molecules in the gas. These
colliding particles could obtain their energy
either from absorbing stellar UV photons with energies
less than 11.2~eV (more energetic photons cannot be present, they would
dissociate H$_2$), or from shocks. In order to investigate this point
further, we examined the 12\,000~K post-AGB star
model from run 4 with graphite dust and no dust formation in the
post-AGB wind (van Hoof et al.\ \cite{vHoof97}).  This model should
give a fair approximation for the physical conditions in
IRAS~16594$-$4656.  Using the distance estimate from
Sect.~\ref{distance} and the known angular dimensions of
IRAS~16594$-$4656, we noticed that the radial sizes in this model were
approximately a factor of two too large.  Hence we reduced all radii
by a factor of two and increased all densities by a factor of four. We
then recalculated the model using Cloudy 96 beta 4 (Ferland
\cite{Ferland02}). This showed that in regions where molecular
hydrogen is abundant, the electron temperature is 210~K or lower. This
value is significantly lower than the rotational temperature of 1440~K
which we derived above. It is therefore very unlikely that H$_2$ is
thermally excited by UV heated gas.

\section{Discussion}
\label{disc}

Based on the evidence presented in the previous section, we can conclude that
the H$_2$ emission is collisionally excited by shocks in IRAS~16594-4656.

Molecular shocks generally can be described as ``C-shocks'' or as
``J-shocks'', depending on their shock velocity $v_{\rm s}$, the
ambient ionization fraction and the strength of the component of the
ambient magnetic field perpendicular to $v_{\rm s}$ (Draine
\cite{draine80}).  A J-shock is a shock where the heat deposition
length is short compared to the cooling length. J-shocks are
sufficiently powerful to dissociate molecules.  C-shocks occur at
relatively low velocities and moderate to low ionizations and depend
upon the presence of a magnetic field. In C-shocks, gas is accelerated
and heated by collisions between charged particles and neutral
particles.  C-shocks cool mainly through molecular material and grains.  In
C-shocks the radiation is emitted as the gas is being heated in the
shock front; in J-shocks it is emitted after the impulsive heating
event, downstream behind the shock front.

The H$_2$ emission in IRAS~16594$-$4656 appears strong: much stronger
than Br$\gamma$ for instance.  Molecular lines produced by collisional
excitation of H$_2$ will be strongest if the collisions are not
energetic enough to dissociate H$_2$ and lower its abundance
(Hollenbach \& McKee \cite{hollenbach89}). Such conditions exist in
C-shocks because the gas is heated gradually and remains molecular.
Hence large columns of material with temperatures between 1000~K and
3000~K can be produced which generate copious amounts of H$_2$
emission.  Therefore the strong H$_2$ emission in IRAS~16594$-$4656 argues 
in favor of a C-shock.

We compared our line ratios of the H$_2$ emission with the intensities
of ro-vibrational transitions of H$_2$ from C-shock models by Le Bourlot
et al. (\cite{Bourlot02}). Our values are consistent with a 20 --
30~km\,s$^{-1}$ C-shock propagating in 10$^3$~cm$^{-3}$ material. They
are not consistent with any of their other models where the C-shocks
have either a higher velocity or impinge on higher density gas.

However, C-type shocks depend upon the presence of a magnetic field.
There is no direct evidence for the presence of a magnetic field in
IRAS~16594-4656. 
Su et al.\ (\cite{su03}) observed 10\% polarization close to the central
star, which makes it likely that the EDE harbors one.
Direct detection of magnetic fields is very difficult, and has only been
attempted a handful of times. Nevertheless, a detection has been
achieved in four post-AGB stars and PNe.  Miranda et al.\ 
(\cite{miranda01}) detected a toroidal magnetic field in the EDE of
the young PN K3-35. Greaves (\cite{Greaves02}) detected a toroidal
magnetic field around NGC~7027 with an effective scale of 5000~AU and
correctly oriented to collimate the observed bipolar winds, and of a
dimension similar to the base of the outflow.  They also detected a
magnetic field in CRL~2688.  Bains et al.\ (\cite{Bains03}) measured
for the first time the magnetic field strength in a post-AGB star.
They found $B=+$4.6~mG at 1612~MHz and $B=+$2.5~mG at 1667 MH in
OH~17.7-2.0.

The lack of significant ionization (no radio emission or lines typical
of H\,{\sc ii} regions have been detected) is consistent with the 
prerequisites of a C-type shock. 

However, metastable lines such as [O\,{\sc i}] 6300~\AA\ and [C\,{\sc
  i}] 9850~\AA\ have been detected.  These are only produced in
J-shocks.  C-shocks, which have large molecular abundances, tend to
rapidly convert any pre-existing atomic oxygen into molecular form
(Hollenbach \& McKee \cite{hollenbach89}).  The observation of ionic
lines such as [Fe\,{\sc ii}] are also a strong indication of J-shocks,
if one has eliminated the possibility of an H\,{\sc ii} region as the
origin. Few ions exist in C-shocks, which only form when the shocked
gas is largely neutral and molecular.

Another indication in favor of the presence of a J-shock is the wind
speed of 126~km\,s$^{-1}$ derived from the H$\alpha$ P-Cygni profile
(Van de Steene et al.\ \cite{VdSteene00b}). This velocity is too high
for a C-shock, but consistent with velocities for a J-type shock
(Hollenbach \& McKee \cite{hollenbach89}).  

[C\,{\sc i}] 9850~\AA, [N\,{\sc i}] 1.040~$\mu$m, and [Fe\,{\sc ii}]
1.257~\& 1.644~$\mu$m are all predicted to be strong near-infrared
emission lines in J-type shocks, while in the optical [N\,{\sc ii}]
6548~\& 6584~\AA, and [S\,{\sc ii}] 6716~\& 6731~\AA\ would be strong.
Comparing the detected near-infrared line fluxes to the model
predictions presented in Fig.~8 of Hollenbach \& McKee
(\cite{hollenbach89}), we find that they are compatible with an
80~km\,s$^{-1}$ J-shock impinging upon $10^3$~cm$^{-3}$ material, or
alternatively a 70~km\,s$^{-1}$ shock impinging upon $10^4$~cm$^{-3}$
material. This would be consistent with a 126~km\,s$^{-1}$ wind
colliding against the side walls of an hourglass nebula at an angle of
approximately 45\degr\ w.r.t.\ the normal, assuming that the side
walls are expanding at a rate of 16~km\,s$^{-1}$ (Loup et al.\ 
\cite{Loup90}). This value for the impact angle is in fair agreement
with the wide opening angle of the lobes seen in the image in Su et
al. (\cite{su01}).

In the optical (Reyniers, private communication) and infrared
spectra no forbidden lines from ionized species
other than [Fe\,{\sc ii}] were detected.
A weak detection of [N\,{\sc ii}] 6548~\& 6584~\AA\ would be
expected in the optical spectrum according to the $10^3$~cm$^{-3}$
model of Hollenbach \& McKee (\cite{hollenbach89}).  For a given shock
velocity, the ionized states will become relatively weaker for higher
densities since the overall degree of ionization will shift downwards.
Taken at face value this could point to a relatively high density (at
least $10^4$~cm$^{-3}$) at the edge of the lobes.  However, according
to the $10^4$~cm$^{-3}$ model, a weak detection of [N\,{\sc i}]
1.040~$\mu$m would be expected, which was not observed in our infrared
spectrum either.  The critical density of this line at $T_{\rm
  e}=10^4$~K is approximately 10$^7$~cm$^{-3}$, using data from
Zeippen (\cite{zeippen82}) and Dopita et al.\ (\cite{dopita76}). This
value is larger than for any of the other lines listed above, ruling
out an extremely high density and collisional de-excitation of the
line as the cause of the non-detection.  Consequently the
non-detections of [N\,{\sc i}] 1.040~$\mu$m and [N\,{\sc ii}] 6548~\&
6584~\AA\ seem to contradict each other and could be indicative that
IRAS~16594$-$4656 has a low nitrogen abundance.  Bipolar PNe with high
C/O and low N/O ratios are known (e.g., J~900, Perinotto
\cite{perinotto91}), although bipolar PNe are usually associated with
type~I PN, which have high nitrogen abundances.  One should keep in
mind though that the Hollenbach \& McKee (\cite{hollenbach89}) models
use solar abundances.  The carbon-rich chemistry in IRAS~16594$-$4656
could have an important effect on the models since it will alter the
cooling significantly.  Deeper observations are needed to settle this
point.
  
>From the previous discussion it is seems that slower C-type and faster
J-type shocks both are present in IRAS~16594$-$4656.  Because the
ionization potential of neutral iron is 7.87~eV, and the dissociation
energy of H$_2$ is 4.48~eV (Graham et al.\ \cite{Graham87}), in
principle H$_2$ and Fe$^+$ cannot coexist in the same region
in substantial quantities. Hence, the C- and J-shocks must occur in
different regions in the nebula.
 
C-shocks would occur where the shock velocity is low, hence where the
fast wind impinges almost tangentially upon the circumstellar
material.  The H$_2$ emission would originate from this region.
We postulate that the C-shocks occur where the fast wind is funneled
through the EDE of IRAS~16594$-$4656.  This is in agreement with
Guerrero et al.\ (\cite{Guerrero00}) who noted that, for PNe, in all
cases where the emission observed in the H$_2$ line is more than 5
times greater than the Br$\gamma$ emission, the H$_2$ emission is
mainly concentrated in a bright ring, the ``waist'' of the bipolar
nebula.  For IRAS~16594$-$4656 the Br$\gamma$ line is much weaker than
the H$_2$ (1,0)\,S(1) line.  The ratio H$_2$ (1,0)\,S(1) to Br$\gamma$ is
8.4 after correction for extinction. As mentioned above, the EDE is
likely to harbor a magnetic field, which is a prerequisite for a
C-shock.

However, molecules have been known to survive in dense condensations
and cometary knots (e.g., Speck et al. \cite{speck02}). Hence it
cannot be excluded that some H$_2$ emission is present in the lobes,
excited by shocks caused by a molecular outflow impinging on the AGB
envelope (e.g., the Egg Nebula, Cox et al. \cite{Cox00}).  The large
difference seen between the {\it IRSPEC} and {\it SOFI} value of
H$_2$(1,0)\,S(1) may support this.  However, no collimated outflow nor
high density blobs have been observed in IRAS~16594-4656 yet. 

J-shocks occur where the shock velocity is high.  A first possibility
is that the [C\,{\sc i}] and [Fe\,{\sc ii}] line emission originates
in a J-shock along the walls of the bipolar lobes, where the fast wind
impinges with a high transverse velocity. This is in agreement with
the observations of Hora \& Latter (\cite{hora96}) and Welch et al.\ 
(\cite{welch99}) who imaged the shock excited [Fe\,{\sc ii}] line
emission in the young PN Hubble~12.  The shape of this PN resembles a
tilted hourglass. The [Fe\,{\sc ii}] emission is located along the
side walls of the hourglass in this object and its morphology differs
strikingly from that of the H$_2$ emission, which is mostly dominated
by UV excitation.  The geometry of this PN is very similar to
IRAS~16594$-$4656, but Hubble~12 is a PN and hence at a much more
evolved stadium. \\
A second possibility is that the [Fe\,{\sc ii}] emission originates much
closer to the central star, in the post-AGB wind itself. Shock waves
induced by stellar pulsations were proposed to explain the [Fe\,{\sc
ii}] emission in Mira variables (Richter et al.\ \cite{richter03}).
Post-AGB stars are of course hotter than Miras, but they are usually
variable and still pulsating. Fokin et al.\ (\cite{fokin01})
argue that stellar pulsations are forming shocks in the atmosphere of
the post-AGB star HD 56126. Miras are believed to have very high mass
loss rates, but the mass loss rate in post-AGB stars may still be quite
large (up to 10$^{-5}$-10$^{-6}$~M$_\odot$\,yr$^{-1}$ 
,Gauba et al. \cite{Gauba03}), as may be 
indicated by the strong P-Cygni Balmer lines in the optical spectrum of
IRAS~16594$-$4656. Moreover, post-AGB stars have higher wind velocities,
which lead to the same normalized [Fe\,{\sc ii}] peak fluxes at lower
pre-shock densities. \\
A third possibility is that the J-shocks would occur where matter
transferred from a binary companion hits an accretion disk. At this
stage there is no corroborating evidence that either a binary companion
or an accretion disk exists in IRAS 16594-4656.

High resolution images in H$_2$ and in selected metastable lines will
clarify which shock mechanism occurs where, while high spectral and spatial
resolution infrared spectra will clarify the kinematics around the
central star and in the bipolar lobes.

\section{Conclusions}
\label{concl}

We examined the near-infrared spectrum of IRAS~16594$-$4656.  It shows
strong H$_2$ emission lines and some typical metastable shock excited
lines such as [Fe\,{\sc ii}] 1.257~\& 1.644~$\mu$m.  We argue that the
molecular hydrogen emission is mainly collisionally excited. Its
strength indicates that the H$_2$ emission originates in C-type
shocks.  However, the metastable lines, and especially the [Fe\,{\sc
  ii}] emission lines, indicate the presence of J-type shocks.  These
shocks don't usually coexist.  Hence we postulate that the H$_2$
emission originates mainly where the stellar wind is funneled through
an EDE and therefore impinges almost tangentially upon the
circumstellar material. The [Fe\,{\sc ii}] emission either occurs
along the walls of the bipolar lobes where the transverse shock
velocity would be higher, or could originate much closer to the central
star in shocks in the post-AGB wind itself, or possibly in an
accretion disk.

IRAS~16594-4656 is unique in the sense that it allows us to study
shock excited lines which are not compromised by ionization or
fluorescence. Further high resolution near-infrared spectra are
currently being obtained in order
to clarify the kinematics of this object and the shock mechanisms.

\begin{acknowledgements}
  PvH thanks the Engineering and Physical Sciences Research Council of
  the United Kingdom for financial support.  GVdS acknowledges support
  for this research carried out in the framework of project IUAP P5/36
  financed by the Belgian Federal Office for Scientific, Technical and
  Cultural Affairs.  This paper contains atomic line data obtained
  from the Atomic Line List v2.04, available at
  {\verb#http://www.pa.uky.edu/~peter/atomic#}. We thank J. C.
  Weingartner for stimulating discussions.  This research has made use
  of NASA's Astrophysics Data System.  The photoionization code Cloudy
  was used, which is written by Gary Ferland at the University of
  Kentucky.

\end{acknowledgements}


\begin{thebibliography}{}

\bibitem[2003]{Bains03}
Bains, I., Gledhill, T. M., Yates, J. A., \& Richards, A. M. S. 2003, \mnras, 338, 287
\bibitem[1976]{black76}
Black, J. H., \& Dalgarno, A. 1976, \apj, 203, 132
\bibitem[1992]{burton92}
Burton, M. G. 1992, Aust. J. Phys., 45, 463 
\bibitem[1989]{cardelli89}
Cardelli, J. A., Clayton, G. C., \& Mathis, J. S. 1989, \apj, 345, 245
\bibitem[2000]{Cox00}
Cox, P., Lucas, R., Huggins, P. J., Forveille, T., Bachiller, R., Guilloteau, S., Maillard, J. P., \& Omont, A. 2000, \aap, 353, L25
\bibitem[1976]{dopita76}
Dopita, M. A., Mason, D. J., \& Robb, W. D. 1976, \apj, 207, 102
\bibitem[1980]{draine80}
Draine, B. T. 1980, \apj, 241, 1021
\bibitem[2002]{Ferland02}
Ferland, G. J. 2002, University of Kentucky, Physics Department Internal Report
\bibitem[2001]{fokin01}
Fokin, A. B., L{\` e}bre, A., Le Coroller, H., \& Gillet, D. 2001, \aap, 378, 546
\bibitem[2002]{GarciaH02}
Garc\'\i a-Hern\'andez, D. A., Manchado, A., Garc\'\i a-Lario, P., Dom\'\i nguez-Tagle, C., Conway, G. M., \& Prada, F. 2002, \aap, 387, 955
\bibitem[1999]{GarciaL99}
Garc\'\i a-Lario, P., Manchado, A., Ulla, A., \& Manteiga, M. 1999, \apj, 513, 941
\bibitem[2003]{Gauba03}
Gauba, G., Parthasarathy, M., Kumar, B., Yadav, R.K.S., Sagar, R. 2003, \aap, 404, 305
\bibitem[1987]{Graham87}
Graham, J. R., Wright, G. S., \& Longmore, A. J. 1987, \apj, 313, 847
\bibitem[2002]{Greaves02}
Greaves, J. S. 2002, \aap, 392, L1
\bibitem[2000]{Guerrero00}
Guerrero, M. A., Villaver, E., Manchado, A., Garc\'\i a-Lario, P., \& Prada, F. 2000, \apjs, 127, 125
\bibitem[1999]{hauschildt99}
Hauschildt, P. H., Allard, F., \& Baron, E. 1999, \apj, 512, 377
\bibitem[1991]{hoban91}
Hoban, S., Reuter, D. C., Mumma, M. J., \& Storrs, A. D. 1991, \apj, 370, 228 
\bibitem[1989]{hollenbach89}
Hollenbach, D., \& McKee, C. F. 1989, \apj, 342, 306
\bibitem[1996]{hora96}
Hora, J. L., \& Latter, W. B. 1996, \apj, 461, 288
\bibitem[1999]{hora99}
Hora, J. L., Latter, W. B., \& Deutsch, L. K. 1999, \apjs, 124, 195
\bibitem[1999]{Hrivnak99}
Hrivnak, B. J., Kwok, S., \& Su, K. Y. L. 1999, \apj, 524, 849
\bibitem[2001]{Hrivnak01}
Hrivnak, B. J., Kwok, S., \& Su, K. Y. L. 2001, \aj, 121, 2775 
\bibitem[1996]{Kastner96}
Kastner, J. H., Weintraub, D. A., Gatley, I., Merrill, K. M., \& Probst, R. G. 1996, \apj, 462, 777
\bibitem[1994]{kurucz94}
Kurucz, R. L. 1994, Solar abundance model atmospheres for 0, 1, 2, 4, and 8~km\,s$^{-1}$,
Kurucz CD-ROM No.~19, Smithsonian Astrophysical Observatory, Cambridge, Mass.
\bibitem[2002]{Bourlot02}
Le Bourlot, J., Pineau des For\^ets, G., Flower, D. R., \& Cabrit, S. 2002
\mnras, 332, 985
\bibitem[1990]{Loup90}
Loup, C., Forveille, T., Nyman, L. \AA., \& Omont, A. 1990, \aap, 227, L29 
\bibitem[1996]{martin96}
Martin, P. G., Schwarz, D. H., \& Mandy, M. E. 1996,  \apj, 461, 265
\bibitem[1997]{Martini97}
Martini, P., Sellgren, K., \& Hora, J. L. 1997, \apj, 484, 296
\bibitem[2001]{miranda01}
Miranda, L. F., G\'omez, Y., Anglada, G., \& Torrelles, J. M. 2001, \nat, 414, 284
\bibitem[1980]{nk80}
Neckel, Th., \& Klare, G. 1980, \aaps, 42, 251
\bibitem[1991]{perinotto91}
Perinotto, M. 1991, \apjs, 76, 687
\bibitem[2003]{richter03}
Richter, He., Wood, P. R., Woitke, P., Bolick, U., \& Sedlmayr, E. 2003, \aap, 400, 319
\bibitem[2001]{rudy01}
Rudy, R. J., Lynch, D. K., Mazuk, S., Puetter, R. C., \& Dearborn, D. S. P.
2001, \aj, 121, 362
\bibitem[1998]{Schlegel98}
Schlegel, D. J., Finkbeiner, D. P., \& Davis, M. 1998, \apj, 500, 525
\bibitem[1978]{shull78}
Shull, J. M., \& Hollenbach D. J. 1978, \apj, 220, 525
\bibitem[2002]{speck02}
Speck, A. K., Meixner, M., Fong, D., McCullough, P. R., Moser, D. E., \& Ueta, T.
2002, \aj, 123, 346
\bibitem[1990]{sternberg90}
Sternberg, A. 1990, in Molecular Astrophysics, ed. T. W. Hartquist (Cambridge: Cambridge University Press), 384
\bibitem[2000]{su00}
Su, K. Y. L., Kwok, S., \& Hrivnak, B. J. 2000, American Astronomical Society Meeting 197, \#131.03
\bibitem[2001]{su01}
Su, K. Y. L., Hrivnak, B. J., \& Kwok, S. 2001, \aj, 122, 1525
\bibitem[2003]{su03}
Su, K. Y. L., Hrivnak, B. J., Kwok, S., \& Sahai, R. 2003, \aj, in press (astro-ph/0304400)
\bibitem[1993]{VdSteene93}
Van de Steene, G. C., \& Pottasch, S. R. 1993, \aap, 274, 895 
\bibitem[2000a]{VdSteene00a}
Van de Steene, G. C., van Hoof, P. A. M., \& Wood P. R. 2000a, \aap, 362, 984 
\bibitem[2000b]{VdSteene00b}
Van de Steene, G. C., Wood, P. R., \& van Hoof, P. A. M. 2000b,
in ASP Conf. Ser. 199, Asymmetrical Planetary Nebulae II: From Origins to Microstructures, 
ed. J. H. Kastner, N. Soker, \& S. Rappaport, 191 
\bibitem[1997]{vHoof97}
van Hoof, P. A. M., Oudmaijer, R. D., \& Waters, L. B. F. M. 1997, \mnras, 289, 371
\bibitem[1998]{Weintraub98}
Weintraub, D. A., Huard, T., Kastner, J. H., \& Gatley, I. 1998, \apj, 509, 728
\bibitem[1999]{welch99}
Welch, C. A., Frank, A., Pipher, J. L., Forrest, W. J., \& Woodward, C. E.
1999, \apj, 522, L69
\bibitem[1982]{zeippen82}
Zeippen, C. J. 1982, \mnras, 198, 111
\end{thebibliography}
\end{document}